\documentclass[conference]{IEEEtran}
\usepackage{graphicx}
\usepackage{subfigure}
\usepackage{amsmath}
\usepackage{color}
\usepackage{xspace}
\usepackage{url}

\newcommand{\figwidth}[0] {0.31\textwidth}
\newcommand{\CI}[0] {\texttt{CI}\xspace}
\newcommand{\BI}[0] {\texttt{BI}\xspace}
\newcommand{\CAP}[0] {\texttt{CAP}\xspace}

\graphicspath{{figures/}}
\begin{document}

\title{DTLS Performance in Duty-Cycled Networks}

\author{
    \IEEEauthorblockN{
        Mali\v{s}a Vu\v{c}ini\'{c}\IEEEauthorrefmark{3}\IEEEauthorrefmark{1}\IEEEauthorrefmark{5},
        Bernard Tourancheau\IEEEauthorrefmark{1},
        Thomas Watteyne\IEEEauthorrefmark{4},\\
        Franck Rousseau\IEEEauthorrefmark{1},
        Andrzej Duda\IEEEauthorrefmark{1},
        Roberto Guizzetti\IEEEauthorrefmark{5},
        and Laurent Damon\IEEEauthorrefmark{5}
    }
    \IEEEauthorblockA{
        \IEEEauthorrefmark{3}BSAC, University of California, Berkeley, CA, USA.
    }
    \IEEEauthorblockA{ 
        \IEEEauthorrefmark{1}Grenoble Alps University, CNRS Grenoble Informatics
        Laboratory UMR 5217, Grenoble, France.
    }
    \IEEEauthorblockA{
        \IEEEauthorrefmark{5}STMicroelectronics, Crolles, France.
    }
    \IEEEauthorblockA{ 
        \IEEEauthorrefmark{4}Inria, EVA Team, Rocquencourt, France.\\
        Email: firstname.lastname@\{eecs.berkeley.edu, imag.fr, st.com, inria.fr\} 
    }
}

\maketitle

\begin{abstract}
The Datagram Transport Layer Security (DTLS) protocol is the IETF standard for securing the Internet of Things.
The Constrained Application Protocol, ZigBee IP, and Lightweight Machine-to-Machine (LWM2M) mandate its use for securing application traffic.
There has been much debate in both the standardization and research communities on the applicability of DTLS to constrained environments.
The main concerns are the communication overhead and latency of the DTLS handshake, and the memory footprint of a DTLS implementation.
This paper provides a thorough performance evaluation of DTLS in different duty-cycled networks through real-world experimentation, emulation and analysis.
In particular, we measure the duration of the DTLS handshake when using three
duty cycling link-layer protocols: preamble-sampling, the IEEE\,802.15.4 beacon-enabled mode and the IEEE\,802.15.4e Time Slotted Channel Hopping mode.
The reported results demonstrate surprisingly poor performance of DTLS in radio duty-cycled networks.
Because a DTLS client and a server exchange more than 10 signaling packets, the DTLS
handshake takes between a handful of seconds and several tens of seconds, with
similar results for different duty cycling protocols.
Moreover, because of their limited memory, typical constrained nodes can only
maintain 3-5 simultaneous DTLS sessions, which highlights the need for using DTLS
parsimoniously.
\end{abstract}

\section{Introduction}
\label{sec:intro}

Secure Sockets Layer (SSL) and its successor Transport Layer Security
(TLS)~\cite{rfc5246} are fundamental cryptographic protocols supporting 
secure communication over the Internet. 
With the advent of the Internet of Things (IoT) and its applications, we face
the problem of supporting communication security for IoT devices that present
stringent energy, memory, and CPU constraints. 
Datagram TLS (DTLS)~\cite{rfc6347} is a version of TLS running over UDP used by
the Constrained Application Protocol (CoAP)~\cite{rfc7252}, ZigBee IP and
Lightweight Machine to Machine (LWM2M) (other standards for constrained IoT
networks) to secure IoT application traffic.

Yet, running DTLS on constrained IoT devices is challenging, in particular
because of the amount of traffic needed to establish a DTLS session and of the memory footprint of a DTLS implementation~\cite{hummen14delegation,vucinic14oscar}.
DTLS benchmarks exist~\cite{granjal12effectiveness,raza13lithe} and focus on memory footprint and message size for different cipher suites.

To achieve long lifetimes up to several years on a battery, IoT
devices use duty cycling link-layer protocols---they follow a sleep/wakeup schedule to
minimize the time their radio transceivers are on, which reduces the energy
consumption.
On the other hand, duty cycling results in higher latency and lower
throughput, which has a direct impact on the DTLS performance.
The goal of this paper is to provide a thorough evaluation of
the DTLS performance on top of representative duty-cycled networks.
%
\begin{figure}
\centering
\includegraphics[width=0.7\columnwidth]{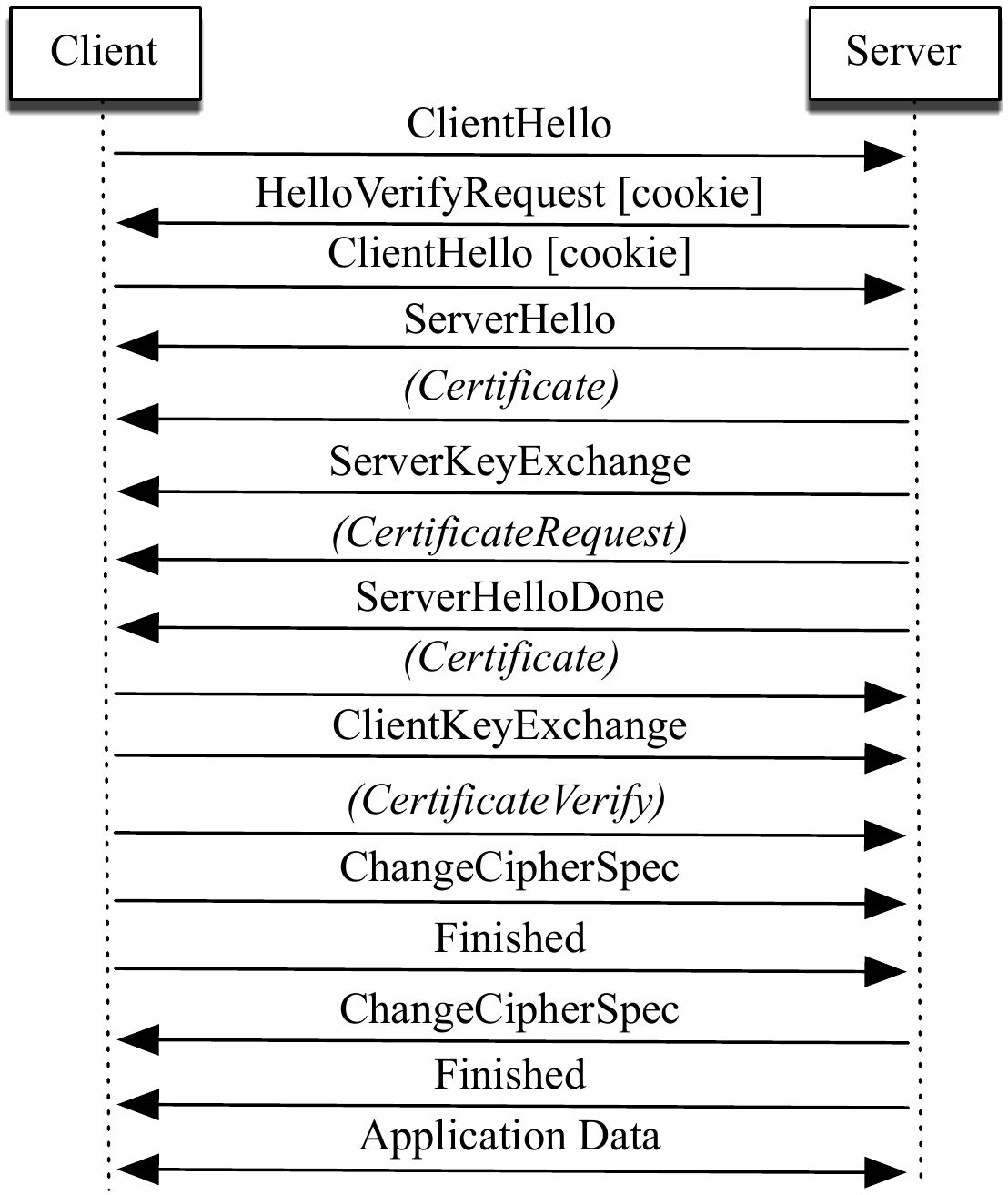}
\caption{Message exchange during a DTLS handshake. Messages in parentheses are not sent for pre-shared key cipher suites.}
\label{handshake_diagram}
\end{figure}
More specifically, the contributions of this paper are the following:
\begin{itemize}
    \item We measure the duration of the DTLS handshake  and energy consumption for the following
      three duty cycling protocols: preamble sampling~\cite{buettner06xmac},
      IEEE\,802.15.4 beacon-enabled mode~\cite{stdieee802154}, and
      IEEE\,802.15.4e Time Slotted Channel Hopping
      (TSCH)~\cite{stdieee802154e}. We use several evaluation methods
      (emulation, measurements on a real-world testbed, an analysis) to obtain
      meaningful results. This part is
      the core of the paper.  
    \item We quantify the impact of the limited memory on the number of simultaneous DTLS sessions a constrained IoT device can maintain.
    \item We show that the probability for a DTLS session establishment to fail
      because the server runs out of memory to hold the associated state can be
      modeled with the Engset loss formula.
\end{itemize}

The remainder of this paper is organized as follows.
Section~\ref{sec:dtls} gives an overview of the DTLS protocol.
Sections~\ref{sec:preamble},~\ref{sec:beacon} and~\ref{sec:tsch} present the
experimental and analytical performance results of DTLS on top of duty cycling protocols.
Section~\ref{sec:memory-constraints} discusses the impact of memory constraints on DTLS.
Section~\ref{sec:related} presents related work and Section~\ref{sec:conclusion} concludes the paper.

\section{Overview of DTLS}
\label{sec:dtls}

Securing application traffic is often achieved by transferring data over a secure channel between the two communicating end-points.
In the network stack, this secure end-to-end channel can be established at the network layer (e.g.~IPsec), the transport layer (e.g.~TLS), or the application layer (e.g.~SSH).
For application development, security at the Transport layer is the most common.
The \textit{de-facto} security standard for the Internet application traffic is Secure Sockets Layer (SSL) and its IETF successor Transport Layer Security (TLS)~\cite{rfc5246}. 
TLS was designed for client-server applications that operate over a reliable transport.
To establish a secure channel, a client and a server first perform the TLS
handshake during which they authenticate each other and derive the symmetric
keys to use during the session.

DTLS~\cite{rfc6347} is an extension of TLS for datagram transport and runs over UDP rather than TCP.
Like TLS, DTLS protects the payload with encryption and authentication.
DTLS records are 8 bytes longer than in TLS, as DTLS adds an explicit 8-byte sequence number.
Stream ciphers, such as RC4, create an inter-record cryptographic context that introduces vulnerabilities with dropped and reordered packets. 
Consequently, DTLS bans the use of stream ciphers and relies on block ciphers for encrypting and authenticating records. 

All messages carried by DTLS are encapsulated within DTLS {\it records} that add a constant 13 byte overhead per datagram.
The Record Layer supports four DTLS upper protocols: 1) {\em Handshake} protocol
establishing a secure authenticated session between two peers, negotiating
algorithms, and the key material; 2) {\em Alert} protocol signaling session
closure or eventual errors; 3) {\em Change Cipher Spec} protocol signaling
modifications to encryption strategies; and 4) {\em Application Data} protocol
carrying application data. 
To deal with Denial of Service (DoS) attacks, the Handshake protocol uses a
stateless {\it cookie} exchange: before the server allocates any resources, the
client needs to resends the {\it cookie} thus proving that the client can
receive messages at a given IP address.

Fig.~\ref{handshake_diagram} shows a message exchange during a DTLS handshake.
Once the client has replayed the stateless {\it cookie} from the server {\tt HelloVerifyRequest} message, the server allocates the necessary resources.
It chooses its preferred cipher from the client cipher set and notifies the
client using a {\tt ServerHello} message.
The messages exchanged during key negotiation depend on the cipher.
When using a pre-shared key, the message containing the server certificate is omitted and the server optionally sends {\tt ServerKeyExchange} indicating to the client which pre-shared key to use.
In this case, the two parties authenticate each other with the common secret
(also used to derive the session keys). 

Because of different application types, DTLS has been used differently in IoT networks and on the traditional Internet.
It is very common for a regular Internet host to establish short-lived TLS
sessions, for example when browsing the Internet {\tt https} URLs.
An IoT device typically periodically reports sensor readings to a server and
therefore uses one long-lived DTLS session, which  is a good thing as constrained IoT networks cannot handle frequent expensive DTLS handshakes, as highlighted in this paper.
However, we are witnessing the emergence of applications in which mobile workers establish short-lived DTLS sessions with individual nodes using hand-held devices, for example for maintenance or drive-by metering~\cite{seitz14usecases}. 
In this context, it is important to understand the limitations of DTLS when duty cycling protocols are used.

Duty cycling is a cornerstone technique for achieving long
lifetimes of IoT devices.
A typical IoT node with an IEEE\,802.15.4 radio will deplete a $2200$\,mAh AA battery in about a week, if the radio is left on continuously (either receiving or transmitting).
State-of-the-art duty cycling protocols reduce the radio duty cycle below 1\%,
thereby extending the device lifetime to several years.
The price of such aggressive duty cycling is an increased network delay and reduced throughput.
In the following sections, we study the effects of duty cycling protocols on the performance of DTLS through emulation, real-world experimentation, and analysis.

\section{DTLS Performance in Duty-Cycled Networks}

\subsection{Preamble Sampling Protocols}
\label{sec:preamble}

\begin{figure*}
\noindent
\centering
\subfigure[DTLS handshake duration, single-hop.]{
    \includegraphics[width=\figwidth]{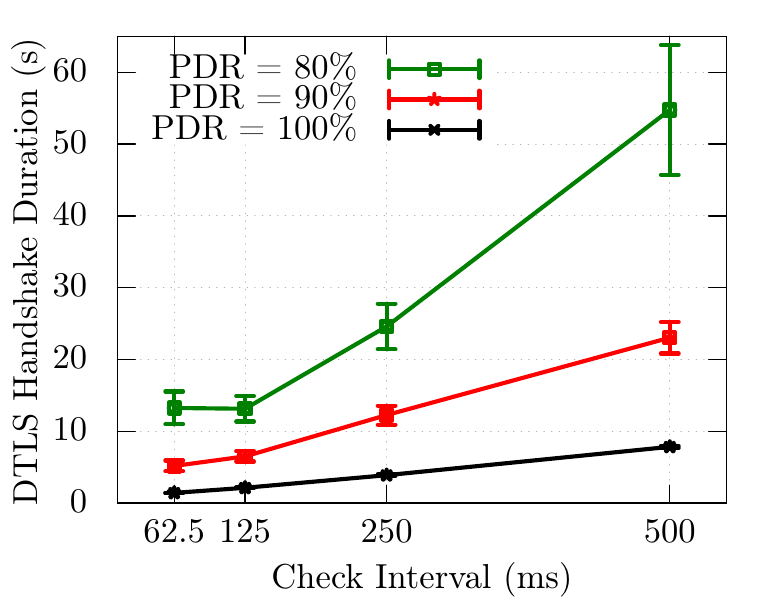}
    \label{wismote_latency_f_bo}
}
\subfigure[DTLS handshake duration, multi-hop (PDR=100\%).]{
    \includegraphics[width=\figwidth]{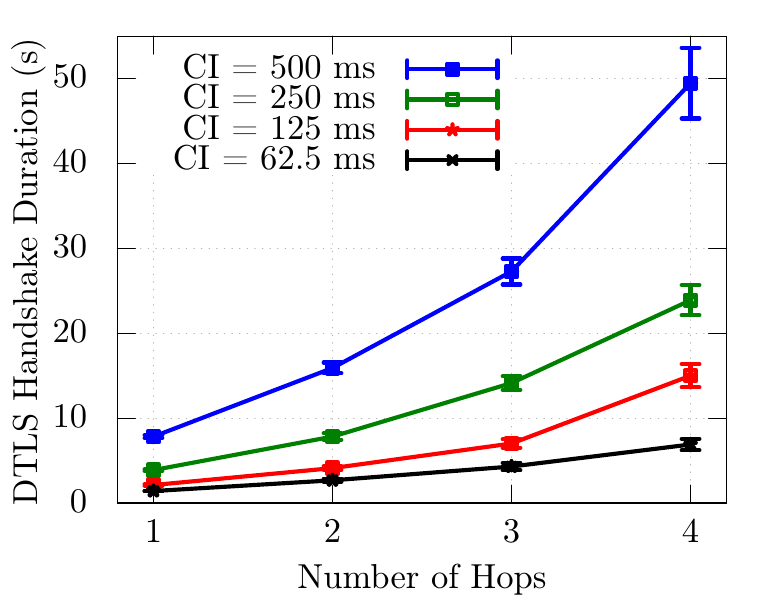}
    \label{wismote_latency_f_hops}
}
\subfigure[Energy, single-hop.]{
    \includegraphics[width=\figwidth]{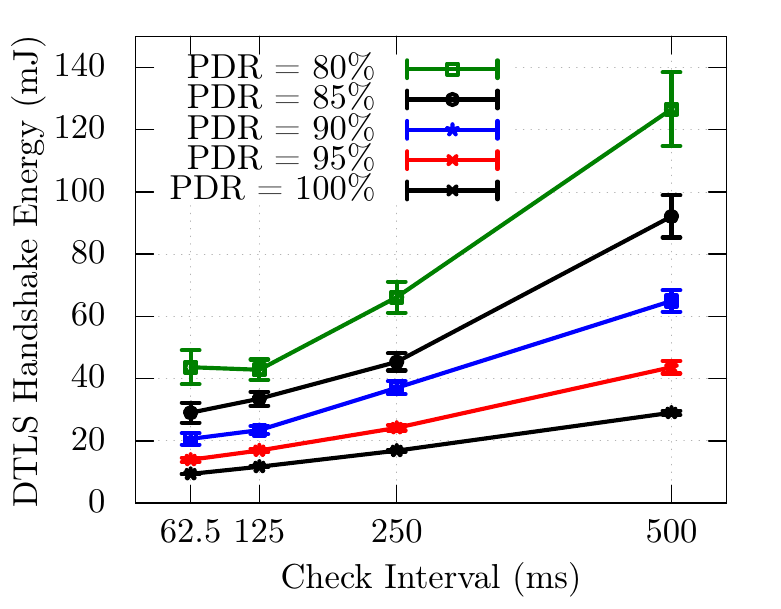}
    \label{wismote_energy_f_ccr}
}
\caption{Cost of a DTLS handshake in preamble sampling protocols.}
\label{preamble_results}
\vspace{-0.5cm} 
\end{figure*} 

To reduce radio idle listening, nodes in preamble sampling protocols periodically wake up to check the state of the wireless medium.
The period at which a node wakes up is called the ``Check Interval'' (\CI); the lower the \CI, the more often nodes check the medium, and the higher their idle radio duty-cycle.
Before sending a data packet, the transmitter repeatedly transmits a special control frame (called \textit{strobe}) for at least the \CI.
If the receiver receives such a frame when checking the medium, it leaves its radio
on until it receives the data frame. In this section, we use
X-MAC~\cite{buettner06xmac,dunkels11contikimac}, arguably the most popular preamble sampling MAC protocol.
X-MAC adds the receiver address in each \textit{strobe}, so only the destination keeps its radio on to receive the data.
All nodes (client, server and relay nodes) use the same \CI value.

We leverage tinyDTLS\footnote{~\url{http://tinydtls.sourceforge.net/}}, an
open-source DTLS implementation and its port to the Contiki operating system~\cite{contiki}. 
We use a pre-shared key cipher suite of DTLS with AES operating in CCM mode with 8-byte long authentication tags ({\tt TLS\_PSK\_WITH\_AES\_128\_CCM\_8}).
We evaluate the performance of this implementation by emulation, using the instruction-level MSP430 emulator MSPSim, and the Contiki Cooja simulator~\cite{lcn06}.
We emulate WiSMote, a popular constrained IoT platform featuring a 16-bit MSP430 micro-controller with $16$\,kB of RAM running at $12$\,MHz, and the IEEE\,802.15.4-compliant CC2520 radio. 
The binary file used for emulation runs on WiSMote hardware.

In the Cooja simulator, we create a linear network of 2 to 5 nodes, depending on
the required number of hops.
The DTLS client (on one end of the linear network) repeatedly performs the handshake with the DTLS server (on the other end).
There is no other traffic in the network. We estimate the energy consumption
using Energest, a Contiki per-component profiling tool that measures the consumption of both the micro-controller and radio.
We average the results over 1000 DTLS handshakes and present with a 95\% confidence interval.

\textbf{Overall results.}
Fig.~\ref{preamble_results} shows the measured average handshake duration and
the energy consumption when DTLS runs on preamble sampling protocols, in the
single/multi-hop case, and for different values of \CI and link
PDR\footnote{Packet Delivery Ratio (PDR) of a link is the percentage of frames
  successfully received at the receiver node. We use the same PDR for all links
  in the emulated network and control its level through the UDG radio model of Cooja.}.
The energy consumption in Fig.~\ref{wismote_energy_f_ccr} is that of the DTLS client (running at $2.8$\,V), during the DTLS handshake.
Although absolute energy values are specific to WiSMote, this platform is representative of hardware commonly deployed today, and the trends in Fig.~\ref{wismote_energy_f_ccr} apply to all platforms.
Overall, a DTLS handshake takes $1$--$50$\,s, with an energy consumption in $10$--$200$\,mJ range.

\textbf{Impact of \CI.}
At PDR=100\%, the DTLS handshake duration and energy consumption grows linearly with \CI.
This is expected, as a larger \CI reduces the rate at which nodes can exchange packets (hence a longer duration).

\textbf{Impact of the number of hops.} 
Similarly, separating DTLS server and client by additional hops increases the duration of the handshake.
For PDR=100\%, the increase is linear (some retransmissions still occur due to the hidden terminal problem); for PDR$<$100\%, the increase is faster as a packet can be lost on each of the hops.

\textbf{Impact of PDR.} 
In any wireless environment, external interference and multi-path fading cause the PDR to be below 100\%.
When a DTLS message is lost, a timeout event occurs at the DTLS layer, which triggers retransmission of DTLS messages.
X-MAC implicitly recovers from lost \textit{strobes}, but does not detect failed receptions of data frames as there are no link-layer acknowledgements.
This means that, when a DTLS packet is lost, the DTLS implementation waits for $2$\,s (the default tinyDTLS timeout value) before resending, causing a longer latency.
Dropping the PDR from 100\% to 90\% roughly triples the handshake duration (Fig.~\ref{wismote_latency_f_bo}) and doubles the energy (Fig.~\ref{wismote_energy_f_ccr}).

\textbf{Energy Consumption.}
Fig.~\ref{wismote_energy_f_ccr} shows how a DTLS handshake consumes more energy
with a larger wakeup interval (longer sleep periods): increasing \CI requires a transmitter to send a longer preamble.
At PDR=100\%, this increase is linear.
To put this energy into perspective, a DTLS handshake cost of $29.05$\,mJ (\CI=$500$\,ms, PDR=$100\%$) represents a consumption 5 orders of magnitude lower than the energy available in a pair of AA batteries.
A single DTLS handshake has hence a negligible effect on the constrained node lifetime.
The cost of completing a single DTLS handshake might be more prohibitive for energy harvesting nodes with small rechargeable batteries,
For example, ST~GreenNet~\cite{varga15greennet} nodes have a $20$\,mAh battery, or $201.6$\,J at $2.8$\,V.
In this context, a single $29.05$\,mJ handshake accounts for $0.0144\%$ of the maximum available energy.

\textbf{Packet overhead.}
Once the DTLS session has been established, DTLS with AES\_CCM\_8 cipher adds 29
bytes to each datagram (including an 8-byte nonce and 8-byte authentication
tag), which  represents $22.8\%$ of the available link-layer payload space (127 bytes).
For the details on byte overhead and possible optimizations, see Raza~\textit{et al.} \cite{raza13lithe}.

\subsection{Beacon-Enabled IEEE\,802.15.4 Networks}
\label{sec:beacon}

Nodes in a beacon-enabled IEEE\,802.15.4 network are organized as a cluster
tree: some nodes are cluster heads (or coordinators), others are leaf nodes.
Cluster heads periodically send beacon frames.
A beacon frame indicates the start of a Contention Active Period (CAP), during
which leaf nodes associated with the emitting cluster head communicate using CSMA/CA medium access.
After the CAP, and before the next beacon, all nodes switch their radio off.
Because beacons are sent periodically, leaf nodes know when to wake up for the next CAP.
The beacon interval (\BI) and the duration of the CAP (\CAP) are tunable, allowing a trade-off between the amount of data that can be exchanged, and the radio duty cycle.

In our experiments, the DTLS client runs on a leaf node.
In the single-hop case, the DTLS server runs on the cluster tree root, otherwise as a leaf node associated with the cluster tree root.
We force a desired topology by tuning the parameters that specify the maximum number of associated cluster heads/leaf nodes such that the association requests are rejected until we obtain the desired topology.
That is, we chain intermediate cluster head nodes between the DTLS client and
the cluster tree root to obtain a linear network from 2 to 5 nodes, depending on
the required number of hops.
These cluster heads do not generate any application traffic; they only transmit periodic beacons and forward packets exchanged between DTLS server, associated with the cluster tree root, and the client.

We implement this protocol on a ST~GreenNet node~\cite{varga15greennet}, an
energy harvesting prototype from STMicroelectronics (ST) based on an ultra low power 32-bit ARM~Cortex-M3 micro-controller (STM32L) with $32$\,kB of RAM and an IEEE\,802.15.4-compliant radio transceiver.
On top of the IEEE\,802.15.4 beacon-enabled mode, the ST~GreenNet nodes run an enhanced
Contiki IPv6/6LoWPAN stack optimized for ultra low power consumption, suitable
for energy harvesting nodes~\cite{varga15greennet}.
We port the tinyDTLS implementation to this stack and modify it to use an AES hardware acceleration block available within the micro-controller.
We run tinyDTLS with exactly the same configuration as in
Section~\ref{sec:preamble} and estimate the energy consumption using Energest
that measures the time spent by  different components of the platform in a given
state (for instance, the time CPU spent in active or low power mode; radio
transceiver in RX or TX mode). 
We then derive energy by multiplying the values by the supply voltage and the current draw values from appropriate data sheets. 

We obtain all the results in this section from measurements on real
hardware. They are
averaged over at least 500 handshakes and presented with a $95\%$ confidence
interval. 

\begin{figure*}
\noindent
\centering
\subfigure[DTLS handshake duration, single-hop.]{
\includegraphics[width=\figwidth]{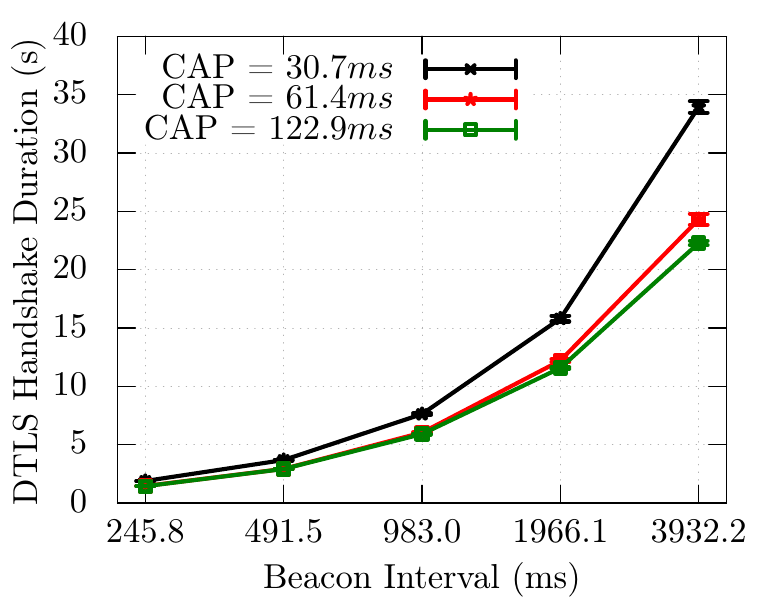}
\label{greennet_latency_f_bo}
}
\subfigure[DTLS handshake duration, multi-hop.]{
\includegraphics[width=\figwidth]{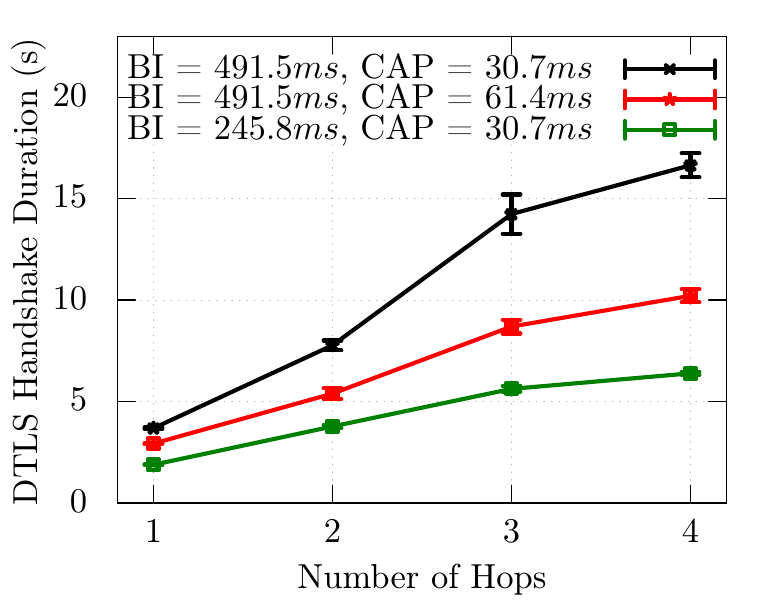}
\label{greennet_latency_f_hops}
}
\subfigure[Energy, single-hop.]{
\includegraphics[width=\figwidth]{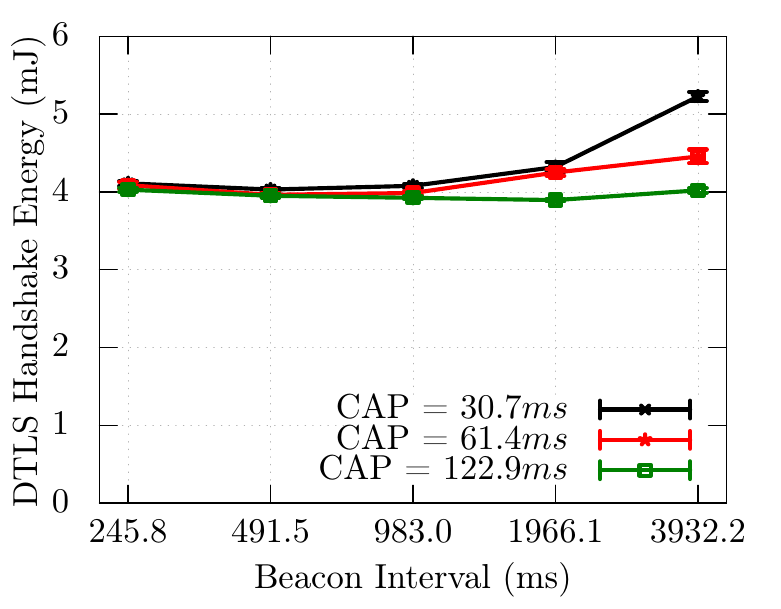}
\label{greennet_energy_f_bo}
}
\caption{Cost of a DTLS handshake in a beacon-enabled  IEEE\,802.15.4 network.}
\label{beacon_results}
\vspace{-0.5cm} 
\end{figure*}

\textbf{Overall results.}
Fig.~\ref{beacon_results} shows the measured average handshake duration and the energy consumption when DTLS runs on an IEEE\,802.15.4 beacon-enabled network, in the single/multi-hop case, and for different values of \BI and \CAP.

\textbf{Impact of \BI.}
Results in the single-hop case (Fig.~\ref{greennet_latency_f_bo}) show how a short \BI shortens the handshake as nodes get more frequent opportunities to transmit.
Similarly, a large \CAP gives nodes a long period to communicate; largest evaluated CAP of $122.9$\,ms yields shortest duration of the handshake.
A larger \CAP increases the throughput between two nodes, which means that a DTLS endpoint (client or server) can send its messages within the same CAP.

\textbf{Impact of the number of hops.} 
Fig.~\ref{greennet_latency_f_hops} presents the measured DTLS handshake duration when DTLS client and server are separated by multiple hops.
As expected, the DTLS handshake duration increases almost linearly with the number of hops.
For values of \BI above $250$\,ms, the successful completion of a DTLS handshake between client and server multiple hops away requires the configuration of a large retransmission timeout value, even when there are no packets lost in the network.
We observe handshake durations from $1.88$\,s to $16.6$\,s.

\textbf{Energy Consumption.}
Fig.~\ref{greennet_energy_f_bo} shows the energy consumed by an ST~GreenNet board running as a DTLS client during a DTLS handshake.
The energy consumption only very slightly increases with \BI, as the energy consumption of a transmitting node in beacon-enabled mode is not a function of the wakeup interval.
Why the energy increases at all with \BI is a consequence of the energy spent by the nodes when sitting idle.

\subsection{IEEE\,802.15.4e TSCH Networks}
\label{sec:tsch}

External interference and multi-path fading severely degrade the quality of a wireless link, both in indoor and outdoor deployments.
The IEEE\,802.15.4e-2012 standard~\cite{stdieee802154e} defines the Time Slotted Channel Hopping (TSCH) mode, which uses ``channel hopping'' to combat external interference and multi-path fading.

In a TSCH network, time is cut into timeslots, each long enough for a transmitter to send a data packet to a receiver, and for the receiver to send back an acknowledgment (ACK) indicating successful reception.
$L$ successive timeslots form a ``slotframe'', which continuously repeats over time.
A communication schedule indicates to each node, for each slot in the slotframe, what to do (transmit, receive or sleep) and on which channel.
The communication schedule can be built in a centralized or distributed fashion.
The scheduler (either a centralized computer or a distributed protocol) builds and maintains the schedule in order to match the link-layer resources (timeslots scheduled between neighbor nodes) to the applications needs (number of packets per second, latency requirements).
We assume that the scheduler schedules a cell (a [timeslot, channel] pair) to only a single pair of nodes, thereby avoiding self-interference in the network.

We derive the expected latency in a TSCH network analytically, and apply it to DTLS.
Let $C$ denote the number of cells scheduled in a slotframe between two nodes in a TSCH network.
We consider that cells are distributed in the TSCH schedule in a uniform fashion, i.e.~the probability for a cell to be assigned to the appropriate timeslot is $1/L$.
Consider a single-hop communication between two nodes; we are interested in
finding the average latency $D$ that includes the time a frame spent in a node queue before its transmission and reception at the destination node.

Let random variable $T$ denote the instant in a slotframe when a frame has been
selected from a node queue for transmission.
We consider $T$ to be uniformly distributed over the slotframe length.
Note that the instant $T$ corresponds to a frame that is either self-originated, or received from a neighbor and to be forwarded.
Let $X_0, X_1, \dots, X_{C-1}$ denote random variables that correspond to the interval from instant $T$ until the beginning of the corresponding cell slot.
The average latency until the beginning of the frame transmission is the expectancy of the random variable $Y = min(X_0, X_1, ..., X_{C-1})$.
Since the slotframe repeats in time, variables $X_0, X_1, \dots, X_{C-1}$ are also uniformly distributed on $(0, L-1)$.
Assuming $L>>C$, the average latency until the beginning of the frame transmission is $L/(C+1)$ timeslots.
Eq.~\eqref{single_hop_delay} expresses the single-hop latency (in timeslots), taking into account the duration of the timeslot during which the frame is transmitted, and the Packet Delivery Ratio $P$ over the link.
\begin{equation}
\label{single_hop_delay}
D_{singlehop} =  (1 + \frac{L}{C+1}) \cdot \frac{1}{P}
\end{equation} 
To extend Eq.~\eqref{single_hop_delay} to the multi-hops case, we take into account the varying number of cells on each link.
Considering a centralized schedule, the total latency over $H$ intermediate hops is the sum of individual hop latencies:
\begin{equation}
\label{multi_hop_delay}
D_{multihop} = \sum_{i=1}^{H}(1 + \frac{L}{C_i+1}) \cdot \frac{1}{P_i}
\end{equation}
We use Eq.~\eqref{single_hop_delay} and Eq.~\eqref{multi_hop_delay} to estimate the average duration of a DTLS handshake for typical TSCH values.
We consider the same scenario as experimentally evaluated in Sections~\ref{sec:preamble} and~\ref{sec:beacon} (10 link-layer frames carrying DTLS messages) and a default timeslot duration of $10ms$.
Tables~\ref{tab:tsch-single-hop-latency} and \ref{tab:tsch-multi-hop-latency} present the estimated DTLS handshake duration for typical values of $L$ and $C$ and ideal Packet Delivery Ratio.

We compared analytical results of TSCH with experimental results of \textit{beacon-enabled} mode in order to find scenarios where they perform similarly. 
For a slotframe length of 101 timeslots, estimated handshake durations with 1, 2, and 3 dedicated cells in TSCH case, roughly correspond to \textit{beacon-enabled} [\BI$=983$\,ms, \CAP$=61.4$\,ms], [\BI$=491.5$\,ms, \CAP$=30.7$\,ms], [\BI$=491.5$\,ms, \CAP$=61.4$\,ms] configurations, respectively. 

\begin{table}
\centering
\caption{Single-hop DTLS handshake duration in a TSCH network.}
\centering
\def\arraystretch{1.2}
\begin{tabular}{crrr}\hline
\bfseries       & \multicolumn{1}{c}{\textbf{C = 1}} & \multicolumn{1}{c}{\textbf{C = 2}} & \multicolumn{1}{c}{\textbf{C = 3}} \\ 
\hline\hline
       L =  101 &           5.15s &          3.467s &          2.625s \\ \hline
       L = 1001 &          50.15s &         33.467s &         25.125s \\ \hline
\end{tabular}
\vspace{-0.5cm}
\label{tab:tsch-single-hop-latency}
\end{table}

\begin{table}
\centering
\caption{Multi-hop DTLS handshake duration in a TSCH network (C=1).}
\centering
\def\arraystretch{1.2}
\begin{tabular}{crrr}\hline
\bfseries       & \multicolumn{1}{c}{\textbf{H = 2}} & \multicolumn{1}{c}{\textbf{H = 3}} & \multicolumn{1}{c}{\textbf{H = 4}} \\ 
\hline\hline
       L =  101 &           10.3s &          15.45s &          20.6s \\ \hline
       L = 1001 &          100.3s &         150.45s &         200.6s \\ \hline
\end{tabular}
\vspace{-0.57cm}
\label{tab:tsch-multi-hop-latency}
\end{table}

\vspace{-0.15cm}
\section{The Impact of Memory Constraints}
\label{sec:memory-constraints}

We have so far focused on the communication aspects of DTLS in duty-cycled networks.
As the role of a DTLS server is often assumed by a constrained device, this section focuses on the effect of memory limitations on DTLS session management.
RFC\,7228~\cite{rfc7228} defines three classes of constrained devices: Class~0 ($\ll10$\,kB RAM, $\ll100$\,kB flash), Class~1 ($\sim10$\,kB RAM, $\sim100$\,kB flash), Class~2 ($\sim50$\,kB RAM, $\sim250$\,kB flash).
According to this classification, WiSMote platform is a Class~1 device, while ST~GreenNet used in Section~\ref{sec:beacon} falls in-between Classes 1 and 2.

Because of the way memory is allocated with embedded processors, a typical implementation statically allocates a number of DTLS ``session slots'', limiting the number of sessions simultaneously open.
The memory footprint for a single DTLS session depends on the cipher suite and key length.
The session state includes the IPv6 address and a port number of the
communicating peer, its role (i.e.~client or server), DTLS engine state, master
secret, derived keys and other implementation specific variables.
As an example, the tinyDTLS implementation uses $\sim400$\,B of RAM per
pre-shared key session, depending on the data type sizes used by different compilers, memory alignment and hardware architecture.
That said, the operating system and the networking stack account for the most of the available memory.
For instance, with $16$\,kB of RAM available on WiSMote nodes, we could only fit 3 DTLS session slots together with the full ContikiOS and the IPv6 networking stack including tinyDTLS, and a simple application for evaluation purposes.

We therefore want to determine the probability $P_{B}$ that a DTLS client attempts to establish a session with a DTLS server where all DTLS session slots are already occupied.
We call $R$ the number of DTLS session slots available at a server.
Let $N$ denote the number of clients interested in establishing a session with the given server.
We model the individual client rate with the exponential distribution of parameter $\lambda$, and the duration of each established session with parameter $\mu$.
Generated traffic in Erlang by each client is then $\rho = \lambda/\mu$.
Under these assumptions, the blocking probability $P_B$ of a DTLS server is simply the blocking probability of a $M/M/R/R/N$ queue, i.e. the Engset loss formula.

For instance, if we consider $R=3$ (the case observed with WiSMote), $N=5$ DTLS clients with $\rho=0.5$, the blocking probability of a request is $\sim17\%$. 
A DTLS server may implement different strategies for handling such requests.
It may discard them or decide to close one of the open sessions in order to accommodate the newly arriving one.
In the latter case, performance depends on the session closure policy, i.e.~the appropriate ``cache'' replacement algorithm. 

\vspace{-0.1cm}
\section{Related Work}
\label{sec:related}

Performance and applicability of DTLS to constrained environments has been a controversial issue for the IoT research and standardization communities.
Many authors studied DTLS in this context
and proposed different optimization techniques~\cite{granjal12effectiveness, raza13lithe,
hummen13viable, hummen14delegation, capossele15security}. 
Granjal \emph{et al.}~\cite{granjal12effectiveness} performed a
comparative study on memory footprints, computational time, and the required
energy between the IPsec protocol and DTLS using different cryptographic
suites.
They showed similar performance of IPsec and DTLS, except when DTLS is additionally used to exchange keys with the Elliptic Curve Diffie-Hellman exchange.

Hummen~\textit{et al.}~\cite{hummen13viable,hummen14delegation} proposed different techniques to lower the impact of the DTLS handshake on constrained devices, certificate pre-validation at the network gateway and handshake delegation to the ``delegation server''.
Raza~\textit{et al.}~\cite{raza13lithe} focused on reducing the per-datagram overhead and proposed a 6LoWPAN DTLS compression scheme.
In our previous work, we studied the benefits of a stateless security architecture called OSCAR~\cite{vucinic14oscar}, based on DTLS, in order to support group communication and caching.

The recent work of Capossele~\textit{et al.}~\cite{capossele15security} explores
the idea of abstracting the DTLS handshake as a CoAP resource and implementing
the handshake procedure using CoAP methods. The advantage of this approach is
that DTLS can leverage the reliability of confirmable CoAP messages, as well as
the blockwise transfer for large messages. The drawback, however, is lost backward compatibility with the existing Internet
infrastructure.

Finally, Kumar~\textit{et al.}~\cite{kumar14hitchhiker} summarized DTLS memory requirements, high level communication overhead (in terms of number of messages), code size for different DTLS and cryptographic functions. 

\vspace{-0.2cm}
\section{Conclusion}
\label{sec:conclusion}

The results of our thorough evaluation reported in this paper demonstrate surprisingly poor performance of DTLS in radio duty-cycled networks.
Experiments with preamble sampling protocols show that the total duration of a DTLS handshake can be more than $50$\,s, depending on the Check Interval. 
In the case of the IEEE\,802.15.4 beacon-enabled mode, we measure durations up to $35$\,s with the largest used Beacon Interval of $\sim4$\,s.
Handshake duration increases linearly with the number of hops for both preamble-sampling and IEEE\,802.15.4 beacon-enabled networks. 
We also derived the analytic expression for latency in TSCH networks and applied it to estimate the duration of DTLS handshake.
For instance, in a typical TSCH network with 101 timeslots per slotframe, where
DTLS server and client are radio neighbors and have a single dedicated cell for
communication, it takes $5.5$\,s on the average  to complete the handshake.
This value decreases to $2.6$\,s for 3 dedicated cells.

Our results show that using DTLS is acceptable for applications for which a DTLS handshake is performed a limited number of times during the constrained node lifetime.
For scenarios that require multiple DTLS clients per DTLS server with constrained resources, we study the blocking probability and show that it corresponds to the Engset loss formula.
Applications expecting a large number of clients per DTLS server should cautiously weight the benefit of its use, against security solutions at the Network or Application layer.

\bibliographystyle{IEEEtran} 
\bibliography{dtls_performance_analysis}

\end{document}